\documentclass[conference]{IEEEtran}
\usepackage{makecell,booktabs}
\usepackage{amsmath,amssymb,amsfonts}
\usepackage{algorithm,algorithmicx,algpseudocode}
\usepackage{graphicx,subfigure}
\usepackage{url}
\usepackage{balance}
\usepackage{textcomp}
\usepackage{xcolor}

\IEEEoverridecommandlockouts

\def\BibTeX{{\rm B\kern-.05em{\sc i\kern-.025em b}\kern-.08em
    T\kern-.1667em\lower.7ex\hbox{E}\kern-.125emX}}
\begin{document}

\title{\huge Digital Twin-assisted Reinforcement Learning for Resource-aware Microservice Offloading in Edge Computing}

\author{ Xiangchun~Chen,~Jiannong~Cao,~Zhixuan~Liang,~Yuvraj Sahni, Mingjin Zhang\\
The Hong Kong Polytechnic University, Hong Kong, China.\\
\{csxcchen, csjcao, cszliang\}@comp.polyu.edu.hk, yuvraj-comp.sahni@polyu.edu.hk, csmzhang@comp.polyu.edu.hk
}

\maketitle

\begin{abstract}


Collaborative edge computing (CEC) has emerged as a promising paradigm, enabling edge nodes to collaborate and execute microservices from end devices. Microservice offloading, a fundamentally important problem, decides when and where microservices are executed upon the arrival of services. However, the dynamic nature of the real-world CEC environment often leads to inefficient microservice offloading strategies, resulting in underutilized resources and network congestion. To address this challenge, we formulate an online joint microservice offloading and bandwidth allocation problem, JMOBA, to minimize the average completion time of services. In this paper, we introduce a novel microservice offloading algorithm, DTDRLMO, which leverages deep reinforcement learning (DRL) and digital twin technology. Specifically, we employ digital twin techniques to predict and adapt to changing edge node loads and network conditions of CEC in real-time. Furthermore, this approach enables the generation of an efficient offloading plan, selecting the most suitable edge node for each microservice. Simulation results on real-world and synthetic datasets demonstrate that DTDRLMO outperforms heuristic and learning-based methods in average service completion time.

\end{abstract}


\section{Introduction}\label{sec:introduction}
The rapid proliferation of the Internet of Things (IoT) devices, with enhanced computation and communication capacities, has given rise to innovative applications like autonomous vehicles, smart healthcare, and metaverse \cite{khan2020edge, xu2022full}. These complex applications involve numerous dependent components and vast amounts of data, necessitating low-latency services that centralized cloud computing struggles to provide. Collaborative edge computing (CEC) has emerged as a promising solution to support these applications \cite{sahni2020multi}. The trend of replacing monolithic applications with loosely-coupled, lightweight microservices is on the rise, given their elasticity and flexibility compared to traditional virtual machines \cite{zhang2022ents}. Microservices offer distinct benefits over their monolithic predecessors, including friendly cross-team development and ease of deployment.

Nevertheless, edge nodes usually operate with limited resources. In critical situations, they may lack the necessary resources to process and transfer heterogeneous application microservices to edge nodes in real-time \cite{zhang2022eaas}. Factors such as high network load, shared radio access network, and an increasing need for network bandwidth, can impede data transmission in edge computing, further straining the already limited resources of edge nodes. Moreover, the resource demands of edge nodes fluctuate over time, indicating a characteristic known as resource-agnostic property. Understanding this dynamic is vital for ensuring equitable resource allocation among competing edge nodes.

Microservices offloading, a key enabling technology in CEC, involves transferring resource-intensive computational microservices from end devices to resource-rich edge nodes. During microservice offloading, the optimal edge node for offloading each microservice is determined based on factors like microservice information, dynamic edge node loads, and network conditions. Data transfer time significantly impacts service completion time, due to microservices require access to data distributed across the network. However, existing works have not adequately addressed joint task offloading and network flow scheduling, resulting in flow congestion and suboptimal performance \cite{tang2020deep,wang2021dependent,wang2021computation}. Some existing works are either limited to data centers or cloud edge computing or offline, rendering them unsuitable for real-time offloading decisions in rapidly changing networks \cite{tran2018joint, sahni2020multihop, sahni2020multi}.


The emerging digital twin technique, which captures system dynamics and provides future insights by integrating data analytics and machine learning methods \cite{khan2022digital}, holds the potential for addressing dynamic resource availability issues. Despite its success in areas like industrial maintenance \cite{mi2021prediction}, aeronautics and astronautics \cite{tao2018digital}, and task offloading in edge computing \cite{liu2021digital}, the integration of digital twin information in microservice offloading and resource allocation remains relatively unexplored. 

To fill this gap, this paper focuses on digital twin-assisted joint microservice offloading and bandwidth allocation in CEC, where the digital twin technique is utilized to predict resource availability. We formulate an online joint microservice offloading and bandwidth allocation (JMOBA) problem to minimize the average service completion time. We address three main challenges. First, how to make optimal decisions considering microservice offloading and network flow scheduling jointly? Second, how do we effectively deal with dynamic resource availability issues of the online CEC environment? Finally, how can digital twin techniques be employed to simulate and adapt to changing edge node loads and network conditions in real-time? To overcome these challenges, we propose a microservice offloading algorithm that integrates digital twin technology with deep reinforcement learning. This algorithm allows us to predict and adapt to real-time changes in loads on edge node and network conditions, adjust offloading decisions based on resource availability, and use a trained DRL model to schedule services efficiently.

The novelty of this work lies in leveraging digital twin techniques to address dynamic resource availability issues in online joint microservice offloading and bandwidth allocation within CEC networks. By employing digital twins, our method can capture the constantly changing resource requirements and network conditions in real-time. This method enables us to make more informed decisions regarding dynamic resource availability, ultimately leading to more efficient microservice offloading and execution. The primary contributions of this paper are as follows:

\begin{itemize}
\item Exploring resource availability dynamics in online joint microservice offloading and bandwidth allocation in CEC. We investigate a novel optimization problem for efficient microservice offloading and bandwidth allocation: JMOBA, which is NP-hard.

\item Building an optimization framework for digital twin-assisted efficient microservice offloading and bandwidth allocation in CEC. The objective is to minimize the average completion time of arrived services.

\item Developing an online microservice offloading algorithm, called DTDRLMO, that utilizes digital twin and deep reinforcement learning techniques. To address dynamic resource availability issues, we design a novel state transition model that leverages digital twin techniques to simulate the state of the CEC environment.

\item Evaluating the proposed algorithm DTDRLMO using both real-world and synthetic datasets through simulations. The results demonstrate that DTDRLMO outperforms state-of-the-art heuristic and learning-based algorithms.
\end{itemize}

The paper is structured as follows. Section \ref{sec:relatedwork} surveys relevant studies. Our system, network models, and problem formulation are in Section \ref{sec:problem-definition}. Section \ref{sec:solution} presents our DTDRLMO approach. The efficacy of our solution is assessed in Section \ref{sec:performance-evaluation}, and Section \ref{sec:conclusion} concludes with our findings and future research work.

\section{Related Work}\label{sec:relatedwork}

\subsection{Task Offloading in Edge Computing}
Joint task offloading and network scheduling in CEC networks has been addressed in several recent studies. Heuristic online methods such as the approach by Meng et al. \cite{meng2019dedas} and OnDisc by Han et al. \cite{han2019ondisc} have been proposed, but their performance is limited due to the dynamic changes in task arrival and network conditions. Deep reinforcement learning (DRL) has been increasingly used to tackle the dynamic issues in task offloading problems in edge computing, with various DQN-based approaches and policy gradient DRL methods \cite{huang2019deep,zhan2020deep,tang2020deep,wang2020fast,wang2021dependent,wang2021computation}. However, these approaches do not jointly consider network flow scheduling, which can lead to network congestion and suboptimal completion time.

\subsection{Microsevice Offloading in Edge Computing}
Several recent studies investigated microservice offloading in edge networks. Samanta et al. \cite{samanta2020dyme} proposed a dynamic scheduling scheme for MEC, optimizing end-user microservices execution while maximizing energy efficiency and QoS. However, they overlooked the overheads of container-based microservices at the edge. Gu et al. \cite{gu2021layer} addressed this by proposing a layer-aware microservice placement and request scheduling scheme, exploiting the layered structure of container images for efficiency. However, the dynamic nature of edge computing environments remains a challenge. 

Only a few studies have addressed the dynamics of edge computing using deep reinforcement learning and digital twin methods. Wang et al. \cite{wang2019delay} and Chen et al. \cite{chen2020iot} utilized DRL to manage dynamic edge environments, but their reliance on prior information and human instruction is a limitation. DT-assisted methods have also been proposed \cite{dai2020deep,liu2021digital,li2022digital}. Dai et al. \cite{dai2020deep} used an asynchronous actor-critic algorithm for stochastic computation offloading and resource allocation. Liu et al. \cite{liu2021digital} employed a combination of the decision tree algorithm and double deep-Q-learning for task offloading to cooperative mobile-edge servers. Furthering DT-assisted methods, Li et al. \cite{li2022digital} studied DT-assisted, Service Function Chain-enabled reliable service provisioning in Mobile Edge Computing networks. 

Contrary to the aforementioned studies, our work explores online joint microservice offloading and bandwidth allocation in CEC with online service arrivals. We further consider the dynamic resource availability issue by leveraging the digital twin technique to provide an accurate prediction of the state of the CEC environment for each microservice.

\section{System Model and Problem Formulation}\label{sec:problem-definition}

\subsection{System Model}
\begin{figure}[t]
	\centering
        \includegraphics[width=\linewidth]{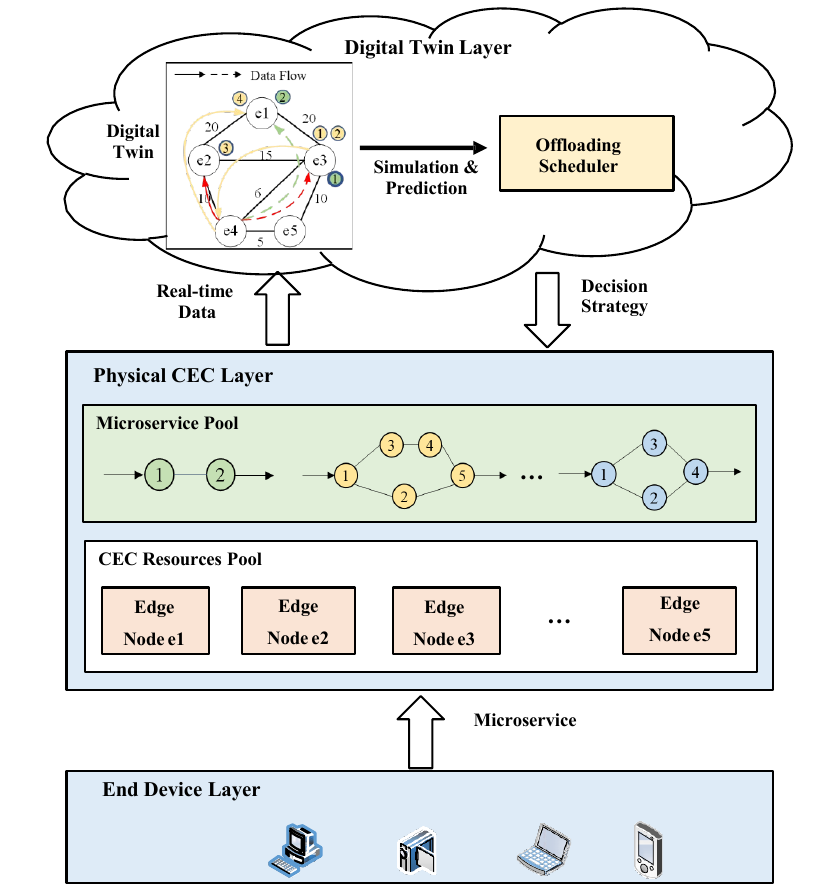}
        \caption{The digital twin-based microservice offloading framework, consisting of the end device layer, physical CEC layer, and digital twin layer.}
        \label{fig_cec}
\end{figure}

Fig.~\ref{fig_cec} presents the architecture of a digital twin-based CEC system. The system comprises the end device layer, physical CEC layer, and digital twin layer. 
\begin{itemize}
\item In the end device layer, the devices submit the application service to edge nodes for processing. 

\item In the physical layer, there are various edge nodes connected via a multi-hop path. The intelligence is dispersed throughout the network by sharing computational resources and data between the edge nodes for collaboratively executing application services. 

\item In the digital twin layer, each edge node and link in the CEC network is mirrored into a digital twin by monitoring and predicting the network status. It also makes offloading decisions about when and where to execute the application services in the CEC resource pool.
\end{itemize}

The CEC system operates in a time-slot manner. The timeline is divided into $K$ time frames (e.g., second), with each frame composed of time slots (e.g., millisecond) of length $T$ with $T\geq 1$. Given a sequence of time slots $\{0, T, \cdots , K*T\}$, we define $t = k*T (k = 0, 1, \cdots)$ as the beginning of each time frame $[t, t+T]$. 

\subsection{Microservice Model}

Assume there are $M$ services submitted to the CEC system. Each service is developed following the microservice principles, which consist of a set of dependent microservices. We use Directed Acyclic Graph (DAG) to model a service. The service set is represented by $CT = \left\{CT_i | 1  \leq  i  \leq  M \right\}$. Each service $i$ is modeled as a DAG, indicated by $CT_i = (T_i, P_{i})$. $T_i$ is the set of dependent microservices in service $i$, $T_i = \left\{j | 1 \leq  j  \leq N_i \right\}$, where $N_i$ is the number of microservices in service $i$. $P_i$ is the set of dependencies among the microservices in service $i$. Each microservice has dependent data with its processor microservices. Each service $i$ is assumed to be generated at an end device and will be offloaded to edge nodes.

Different microservices have different computation and communication resource requirements. The size of the input data packet for microservice $j$ of service $i$ is $D_{i,j}$.  Each microservice $j$ is associated with the computation load $C_{i,j}$. The set of predecessors for microservice $j$ is represented by $Pd_{i,j}$.

\subsection{Digital twin-assisted Collaborative Scheduling}

The network model is represented as a connected graph $G= (V, E)$. Inside, $V = \left\{v | 1  \leq  v  \leq  \alpha \right\}$ is the set of edge nodes, where $\alpha$ is the total number of edge nodes. $E$ is the set of links connecting different edge nodes, $E=\left\{e_{j,k} | j, k \in V \right\}$.

The resource consumption of an edge node $n$ can be expressed by a tuple $⟨R_{i,j}, AR_{n}⟩$, where $R_{i,j}$ is the resource amount required by microservice $j$ in service $i$ and $AR_{n,t}$ is the amount of available resources of edge node $n$. We adopt digital twin techniques to dynamically provide accurate and personalized availability prediction for each edge node.

\subsection{Problem Formulation}
We first consider a snapshot of arrived microservices in the CEC network. Given a CEC network $G = (V, E)$, digital twins run in the remote cloud to provide a reliable prediction of microservices and edge nodes in real-time. The online joint microservice offloading and bandwidth allocation problem is to minimize the average completion time of all services submitted to the CEC resource pool in the time snapshot, while meeting their availability requirements, subject to the computing resource on each edge node in $G$. More specifically, we will decide which edge node to allocate a microservice, when to execute it, and the bandwidth allocated to intermediate data flow among microservices, with digital twin, to dynamically predict the resource availability of each edge node. 

The completion time of a service consists of two parts: the computation cost, which indicates the completion time of a microservice on allocated edge nodes, and the flow transmission cost, which indicates the time spent to receive the dependent input data from processor microservices. The microservice generated at an end device will be offloaded to the local edge node, which is nearest to the end device after its release. So we can ignore the upload time of the service. According to \cite{cheng2018energy}, we can also ignore the effect of the result return time on microservice uploading because the output after microservice processing is much smaller than the input.

\textbf{Computation Cost.} The computation time of edge node $n$ for microservice $i$ is composed of four parts: the time when the service is uploaded to the edge nodes, the waiting time of the microservice at an edge node, the calculation time of the microservice at the edge nodes, and the download time when the result is returned to the user. According to \cite{cheng2018energy}, we can also ignore the effect of the result return time on microservice uploading because the output after microservice processing is much smaller than the input. 

Let $\delta_{i,j,n,t} \in \left\{0,1\right\} $ be a binary decision variable to indicate whether the microservice $j$ in service $i$ is allocated to edge node $n$ at time $t$. Let $W_{i,j}$ represent the waiting time of microservice $j$ in service $i$. $PS^t_{n,t}$ indicates the average processing speed of an edge node $n$ at time $t$. Then, the computation time $PT_{n,i,j}$ of microservice $j$ of service $i$ at edge node $n$ is defined as: 

\begin{equation}
    PT_{n,i,j} = (C_{i,j}/PS^t_{n,t} + W_{i,j})*\delta_{i,j,n,t}
\end{equation}

\textbf{Flow Transmission Cost.} A microservice can not start until it receives all dependent input data. Let $B^t_{f,i,j}$ be a continuous decision variable to indicate the bandwidth allocated to the flow $f$ for microservice $j$ in service $i$ at time $t$. The routing path of flow $f$ is denoted as $P_f$. We assume it is the shortest path between the source node and the destination node. Let $BW^t_{e}$ indicate the remaining bandwidth of edge link $e$ at time $t$. The scale of $B^t_{f,i,j}$ is:

\begin{equation}
0 < B^t_{f,i,j} < \min_{e \in P_f} BW^t_{e}
\end{equation} 

$\epsilon_{f}$ is the waiting time of the flow $f$ to be transmitted. Then, the flow transmission time $BT_{f,i,j}$ of flow $f$ for microservice $j$ is calculated as:

\begin{equation}
    BT_{f,i,j}= D_{i,j} / B^t_{f,i,j} + \epsilon_{f}
\end{equation} 

With the formulation of the computation and flow transmission time, the finish time $FT_{i,j}$ of microservice $j$ in service $i$ can be calculated as:

\begin{equation}
    FT_{i,j} = PT_{n,i,j} + BT_{f,i,j} + \max_{k \in Pd_{i,j}} (FT_{k})
\end{equation}

where $Pd_{i,j}$ represents the set of predecessors of microservice $j$. The finish time $TT_{i}$ of service $i$ is calculated as:

\begin{equation}
    TT_i = \max_{j \in T_i} FT_{i,j} 
\end{equation}

\textbf{Objective.} Our objective is to minimize the average completion time of all services submitted at timeslot $t$. The joint dependent microservice offloading and bandwidth allocation problem can be formulated as follows:

\begin{equation}
    \min_{\delta_{i,j,n,t}, B^t_{f,i,j}, W_{i,j}, \epsilon_{f}} \sum_{i=1}^{M} TT_i /M
\end{equation}

\begin{equation}\label{unique}
     \sum_{n=1}^{\alpha} \delta_{i,j,n,t} = 1, \quad \forall i, j 
\end{equation}
\begin{equation}\label{resource}
    \sum_{i=1}^{M} \sum_{j=1}^{N_{i}} R_{i,j}*\delta_{i,j,n,t} \leq AR_{n,t}
\end{equation}
\begin{equation}\label{bandwidth}
    \sum_{i=1}^{M} \sum_{j=1}^{N_{i}} B^t_{f,i,j} \leq \min_{e \in P_f} BW^t_{e},  \quad    e \in P_{f}
\end{equation}
where Eq.~\eqref{unique} ensures that a microservice can only be allocated to an edge node. Eq.~\eqref{resource} and Eq.~\eqref{bandwidth} show the resource availability constraints. Eq.~\eqref{resource} indicates that the resource consumption of all microservices allocated to node $n$ shall not exceed the available resources of node $n$ at time $t$. Eq.~\eqref{bandwidth} constraints that the allocated bandwidth of a link $k$ shall not exceed the available bandwidth capacity of the link.

\textit{Theorem 1:} The online joint microservice offloading and bandwidth allocation problem in $G = (V,E)$ is NP-hard. 

\textit{Proof:} The online joint microservice offloading and bandwidth allocation problem in $G = (V, E)$ is NP-hard. We will prove that the one-time-slot joint microservice offloading and bandwidth allocation problem is NP-hard by reducing it from the well-known NP-hard problem, the generalized assignment problem (GAP) \cite{cattrysse1992survey}.

The GAP involves packing $N$ items into $M$ bins, where each bin $m$ has a capacity $C(m)$ and packing item $n$ into bin $m$ incurs a cost $cost(n, m)$ and has a size $size(n)$. The objective is to minimize the total cost of packing all items into bins, subject to the bin capacities.

We consider a special case where the network topology $G = (V, E)$ is fully connected, and only microservice offloading is considered. Let $CT$ be the set of microservices. Each edge node $v \in V$ represents a bin associated with a processing capacity $PS_v$, and each microservice $j$ in service $i$ represents an item associated with a size $C_{i,j}$, i.e., the computing resource consumption of microservice $j$. The cost of placing item $i$ into bin $v$ is $cost(i, v)$, i.e., the cost of offloading a microservice $j$ on edge node $v$. We can see that this special problem is equivalent to the minimum-cost GAP. Therefore, the special problem is also NP-hard \cite{oncan2007survey}. Thus, the online joint microservice offloading and bandwidth allocation problem in $G = (V, E)$ is also NP-hard.

\section{Digital Twin-assisted Deep Reinforcement Learning} \label{sec:solution} 
In this section, we model the problem as a Markov decision process (MDP) and propose a novel digital-twin-based deep reinforcement learning approach.

\subsection{Markov Decision Process}

The Markov Decision Process (MDP) is used to model the scheduler in our CEC system. In CEC, the scheduler's action consists of designating an offloading device and allocating bandwidth for each arriving microservice, thereby leading to a state transition in the MDP. The state transition can be represented as $Pr[S_{t+1}=s'| S_t=s, A_t=a]$. $S_t$ denotes the current state, $A_t$ is the action taken by the scheduler, and $S_{t+1}$ represents the next state. The agent is the scheduler itself, which is responsible for determining the action that leads to the most favorable state transition. The ultimate goal of the agent is to maximize the cumulative reward over time, thereby resolves the problem of efficient microservice offloading in the edge network.

\subsection{State Space} 
The state space is defined by the combination of the information of services $CT$ and the network information $NI$. For a given microservice, the state of the CEC system at time $t$ is contingent on the scheduling outcomes of preceding microservices and network flows. 

\begin{equation}
    S = \{s_t | s_t = (CT, NI_t) \}
\end{equation}

\begin{equation}
    CT = \{CT_i | CT_i = (T_i, P_i )\}
\end{equation}

\begin{equation}
    NI_t = \{BW^t_1, ... BW^t_p, AR^t_1, ..., AR^t_\alpha\}
\end{equation}

Specifically, $CT$ includes the information of service DAGs, consisting of the service $CT_i$, a vector $T_i$ of dependent microservices within $CT_i$, and the dependencies $P_i$ among these microservices. $NI_t$ represents the status of the CEC network environment, including the bandwidth conditions of the links and the resource conditions of edge nodes. $p$ is the number of edge links. The remaining bandwidth of edge link $j$ at time $t$ is represented by $BW^t_j$. $\alpha$ is the number of edge nodes. The remaining computation resource (e.g., cpu and memory) of edge nodes $k$ at time $t$ is represented by $AR^t_k$.

\subsection{Action Space} 
The action space comprises the set of potential decisions the agent can make in response to a given state. For our problem, the action space includes offloading decisions for the current microservice and the bandwidth allocation for the corresponding network flow.

\begin{equation}
a_{t} = \left\{ a_{i, t} | a_{i, t} = \left\{ q_{i,t}, \eta_{i,t}, \epsilon_{i,t} \right\} \right\}
\end{equation}

Each action $a_{i, t}$ includes three components: the offloading decision $q_{i,t}$ for each microservice within the current service, the bandwidth $\eta_{i,t}$ allocated to each network flow, and the waiting time $\epsilon_{i,t}$ for each flow.

\begin{figure}[t]
	\centering
        \includegraphics[width=0.83\linewidth]{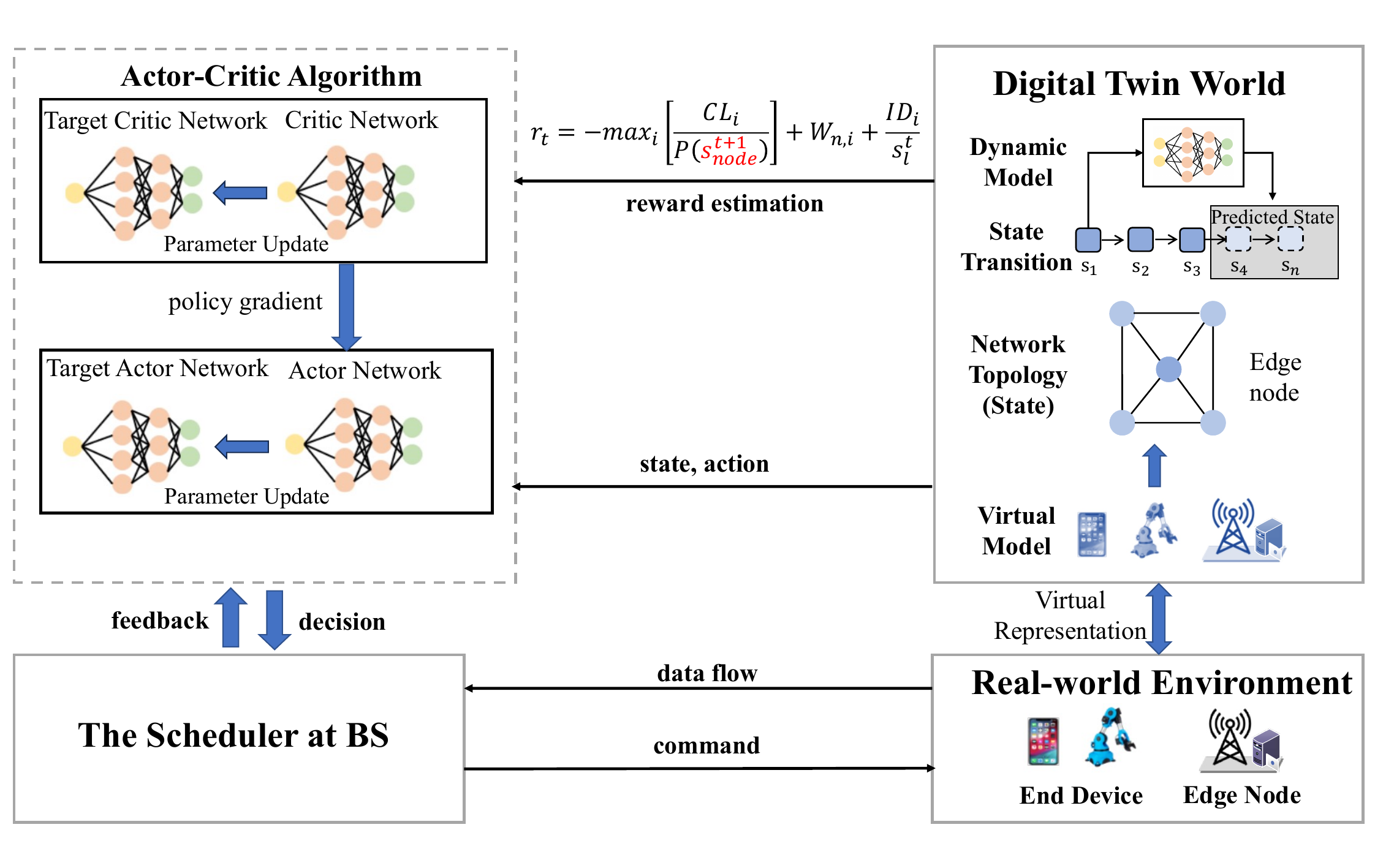}
        \caption{The diagram of digital twin-assisted deep deterministic policy gradient algorithm.}
        \label{fig_ddpg}
\end{figure}

\subsection{Reward Function}
The reward function quantifies the immediate benefit the agent receives after executing an action. In the context of our digital twin-assisted deep deterministic policy gradient (DDPG) offloading algorithm, the reward is derived from the time consumption for completing microservices at the edge nodes. Specifically, the reward $r_{t}$ at time slot $t$ is defined as the negative weighted sum of microservice completion time. This design ensures that the agent is incentivized to minimize the microservice completion time.

\begin{equation}
    r_{t} = -TT_t
\end{equation}

In this equation, $TT_t$ denotes the weighted sum of the completion time of microservices generated at time slot $t$. Each action taken by the agent will result in a new state and associated reward, contributing to the optimization of the offloading process in the digital twin edge network. 

\subsection{Digital Twin-assisted Deep Reinforcement Learning}

\textbf{Overall Training Framework.} Fig.~\ref{fig_ddpg} shows our overall training framework, which includes three key learning components: \textit{the dynamic model, actor network, and critic network}. The dynamic model is aimed at learning the state transition of the environment, while the actor network and critic network are trained to learn the optimal policy. After learning the dynamic model, we use it to predict the next state and leverage this information to estimate the reward.

\textbf{State Transition Model (Environment Dynamic Model).} We use a deep neural network parameterized by $\phi_s$ to approximate the state transition, which maps the state-action tuple to the next state:

\begin{equation}
    \forall s_t \in \mathcal{S}, a_t \in \mathcal{A}, \quad \hat{s}_{t+1}=f_{\phi_s}\left(s_t, a_t\right)
\end{equation}

The model can be optimized through supervised learning by regression using the mean-squared error between the predicted next state and the actual one.

\begin{equation}
    \min _{\phi_s} \frac{1}{|\mathcal{D}|} \sum_{\left\langle s_t, s_{t+1}, a_t, r_t\right\rangle \in \mathcal{D}}\left\|s_{t+1}-\left(s_t+f_s\left(s_t, a_t\right)\right)\right\|^2
\end{equation}

Where $\mathcal{D}$ is the replay buffer and $f_s$ maps the state-action tuple to the difference between the current state and the next state. During each iteration of training, we alternate between model fitting and agent updates using the data collected from the environment interactions. 

\textbf{Estimating the reward using the state model.} Inspired by the model-based reinforcement learning, we use the learned state model to predict the next state so as to estimate the remaining bandwidth and resources. Therefore, the reward function can be further designed as follows:

\begin{equation}
    r_t=-\max _{i \in CT} \max _{j \in CT_j}[\frac{C_{i,j}}{P(s_{node}^{t+1}))}+W_{i,j}+\frac{D_{i,j}}{s_{link}^{t}}]
\end{equation}

Where the $s_{node}^{t+1}$ is the next state predicted from the state transition model.

\begin{algorithm}[t]
	\caption{Digital Twin Modelling Algorithm}
	\label{alg_digital}
	\textbf{Input:} Current state $s$, action $a$, reward $r$, next state $s\_$;
        \begin{algorithmic}[1]
        \Require Number of epochs as $ep$, batch size as $bs$, and learning rate as $lr$.
        \State Initializes the transaction model $t\_model$ and the reward model $r\_model$.
        \State Define the transaction dataset using the input $bs$ and $ba$, and target datasets $bs\_$.
        \State Define the reward dataset using the input $bs\_$ and target datasets $r$.
        \State Train the transaction model $t\_model$ and the reward model $r\_model$ using the defined datasets, epochs, batch size, and learning rate.
        \State The forward propagation $h=f(w_i*X + b_i)$ through the autoencoder is simply applying the activation function (e.g., ReLU, sigmoid, etc.) to the dot product of the input and the weights of the model.
        \State This process is repeated for a specified number of epochs until the model converges, i.e., the loss doesn't decrease significantly with further training. 
 	\end{algorithmic} 
\end{algorithm}

\textbf{Deep Deterministic Policy Gradient.} In accordance with Fig.~\ref{fig_ddpg}, the DDPG model consists of an actor network, a critic network, and an experience pool. The actor network uses a policy-based approach, the critic network is value-based, and the experience pool retains each system state record $(s_t, a_t, r_t, s_{t+1})$. The deterministic strategy $\mu$ is formulated as:

\begin{equation}
 \mu: s_t \rightarrow a_t
\end{equation}

By approximating the strategy with a continuous function parameterized by $\theta$, the deterministic strategy becomes:

\begin{equation}
    \mu_\theta (a_t | s_t) = P(a_t | s_t; \theta)
\end{equation}

The agent's goal is to learn the optimal policy to maximize the progressive reward:

\begin{equation}
    J(\theta)=\mathbb{E}_{s_t \sim \rho^{\mu}, a_t \sim \mu}[r]
\end{equation}

Here, $\rho^{\mu}$ represents the state transition probability given the action distribution $\mu$ and $r$ is the progressive expected rewards.

To address high variance in the policy gradient estimator, the actor-critic network approximates the expected returns with a function $Q_{\omega}(s, a)$, replacing the original return term. The gradient of the objective $J$ is:

\begin{equation}
    \begin{aligned}
    \nabla_{\theta} J(\theta)=\mathbb{E}_{s_t \sim \rho^{\mu}}\left[\nabla_{\theta} \mu_{\theta}(a \mid s_t) \nabla_{a} Q^{\prime}(s_t, a)|_{a=\mu_{\theta}(s_t)} \right]
    \end{aligned}
\end{equation}

A replay buffer stores previous system state transition experiences, which are randomly sampled during training to enhance learning performance. 

To train the critic network parameters, a batch of experiences is randomly drawn from the replay buffer. The training objective is to minimize the difference between $Q(s_t, a_t; w)$ and $Q^{\prime}(s_t, a_t; w^{\prime})$ using the loss function $L$:

\begin{equation}
    L = 1/N \sum_i (y_i - Q(s_t, a_t; w))^2
\end{equation}

\begin{equation}
    y_i = r_i + \gamma Q^{\prime}(s_{i+1}, \mu^{\prime}_{\theta^{\prime}} (s_{i+1}); w^{\prime})
\end{equation}

where $N$ is the number of samples drawn from the experience pool, $y_i$ is the temporal difference target for Q-learning, and $w$ and $w^{\prime}$ represent parameters of the critic network.

As outlined in Algorithm~\ref{alg_ddpg}, the scheduler begins by initializing the replay buffer, critic network, and actor network with weight $w$ and $\theta$ (lines 1-3). After obtaining the environmental state $s_t$, the agent chooses action $a_t$ based on the current policy and exploration noise (lines 5-8). Following action execution and interaction with the environment in the processing pool, the agent receives reward $r_t$, observes the next state $s_{t+1}$ of the environment, and stores the state $\left<s_t, a_t, r_t, s_{t + 1}\right>$ into the replay buffer (lines 9-10). Subsequently, the agent randomly samples a mini-batch experience from the replay buffer and updates the critic network by minimizing the loss (lines 11-13). After each interaction episode with the environment, the agent updates the actor and critic networks using the sampled policy gradient and target network parameters (lines 14-19). 

\begin{algorithm}[t]
	\caption{DTDRLMO Algorithm}
	\label{alg_ddpg}
	\textbf{Input:} Training episode length $Y$, training sample length $T$; DAG microservice number $\beta$, edge node number $\alpha$, edge link number $p$;
	\begin{algorithmic}[1]
	  \State Randomly initialize actor network $A_\theta$ and critic network $c_w$ with weights $\theta$ and $w$, respectively
        \State Initialize target network $A_\theta$ and $c_w$ with weights $w^{\prime}$ and $\theta^{\prime}$, respectively
        \State $w^{\prime} \gets w$, $\theta^{\prime} \gets \theta$
	    \State Initialize replay memory $B$
	    \For {episode $\gets 0$ \textbf{to} $Y$}
	        \State Get environment state $s_t$
	        \State Initialize a random process $N$ for action exploration
    	    \For {$t \gets 1$ \textbf{to} $T$}
                \State Execute each action and obtain the estimated reward $r_t$ and estimated new state $s_{t+1}$ in digital twin work.
                \State Select action $a_t = ( q_t, \eta_t, \epsilon_t )$ with the maximum reward according to the current state, future state, current policy and exploration noise
    	      \State In collaborative edge computing, execute action $a_t$ and obtain the reward $r_t$ and new state $s_{t+1}$
              \State Store state transition $(s_t, a_t, r_t, s_{t+1})$ in pool $B$
              \State Randomly select a minibatch of $N$ transitions $(s_i, a_i, r_i, s_{i+1})$ from $B$.
              \State $y_i \gets r_i + \gamma Q^{\prime} (s_{i+1}, \mu^{\prime}_{\theta^{\prime}} (s_{i+1}); w^{\prime})$
              \State Update the critic network by minimizing the loss $L = 1/N \sum_i (y_i-Q(s_i, a_i;w))^2$
              \State Update the actor network as follows: $\nabla_{\theta_\mu} J \approx 1/N \sum_i Q(s_t, a_t;w)|_{s_l=s_i,a=\mu_\theta (s_i)} \nabla_{\theta_\mu} \mu_\theta (s_t) |_{s_t = s_i}$
              \State $\theta^{\prime}_{\mu^{\prime}} \gets t \theta_\mu + (1-t) \theta^{\prime}_{\mu^{\prime}}$\Comment{Update one of the target networks}
              \State $w^{\prime}_{Q^{\prime}} \gets t w_Q+(1- t) w^{\prime}_{Q^{\prime}}$\Comment{Update the other target network}
            \EndFor
        \EndFor
	\end{algorithmic} 
\end{algorithm}

\textbf{Difference from other existing RL approaches.} Most of the existing RL approaches are model-free based approach, which does not explicitly consider the future state transition. In contrast to the existing model-based RL algorithms, our method uses the state transition model to reshape the reward function, while others only use it to generate experience data.

\section{Performance Evaluation}\label{sec:performance-evaluation}

\subsection{Simulation Settings}

\textbf{Network Model.} Our network model utilizes a stochastic topology generator \cite{TP-toolbox-web} to construct a network topology comprising $8$ edge nodes. The weight (or capacity) of the connection between any two edge nodes signifies the bandwidth of the link. Owing to the heterogeneity in devices, the computational capacity of edge nodes is chosen from a Gaussian distribution with an average of $40$ Mcps (megacycles per second) and a standard deviation of $80\%$. Similarly, the processing power of end devices follows a normal distribution with a mean of $10$ Mcps and a standard deviation of $20\%$. The bandwidth of each edge link connection is also selected from a normal distribution, with an average of $10$ Mbps (megabits per second) and a standard deviation of $80\%$.

\textbf{Synthetic Dataset.} We use a stochastic Directed Acyclic Graph (DAG) generator \cite{sahni2020multihop} to create DAG microservices. Each DAG microservice has properties like microservice id, service id, data size, computational load, release time, and source device. The number of nodes and layers in the DAGs is selected from a Gaussian distribution within $[1, 50]$. The number of edges or dependencies between two layers in the DAG is chosen from a uniform distribution. The microservices are randomly generated at edge nodes. The data size for each microservice is randomly generated from a normal distribution with a mean of $500$ Mbit and a standard deviation of $80\%$. The computational load for each microservice is randomly chosen from a Gaussian distribution with a mean of $500$ kilo clock cycles and a standard deviation of $60\%$. The release time for microservices is randomly chosen using a Poisson distribution. The synthetic dataset is split into training and testing subsets.

\textbf{Real-world Dataset.} Alibaba cluster trace includes DAG details of real-world workload traces \cite{Alidata}. Some modifications are made due to incomplete information in the Alibaba cluster trace to generate DAGs. Computational loads of microservices are obtained by appropriately scaling the value of the difference between "start time" and "end time", multiplied by "plan cpu". The data size of each microservice is randomly chosen from a Gaussian distribution with a mean of $500$ Mbit and a standard deviation of $80\%$. The release time for each microservice is randomly selected using a Poisson distribution. Microservices are randomly spawned at edge nodes. The real-world dataset is split into training and testing subsets.

\textbf{Metric.} In our simulations, we assess the efficacy of the suggested algorithms using the primary metric: Average Completion Time (ACT). ACT is calculated as the total completion time of all services divided by the total number of services. This measure reflects the average duration required for service completion, offering insights into the efficiency of the microservice offloading procedure. 

\begin{figure}[t]
    \centering
    
     \begin{minipage}[t]{0.48\linewidth}
        \centering
        \includegraphics[width = \linewidth]{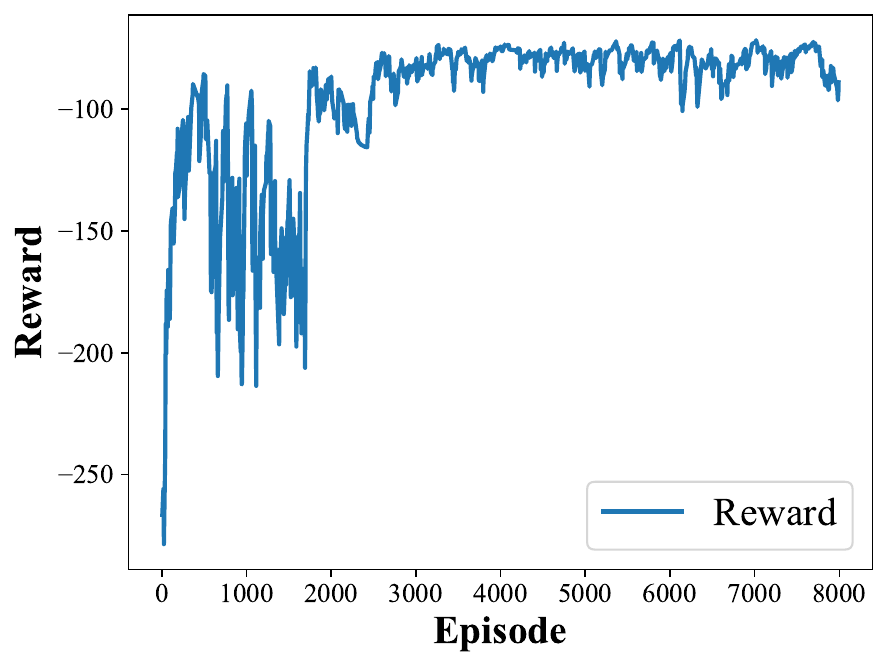}
        \label{fig_reward}
    \end{minipage}
    \begin{minipage}[t]{0.48\linewidth}
        \centering
        \includegraphics[width = \linewidth]{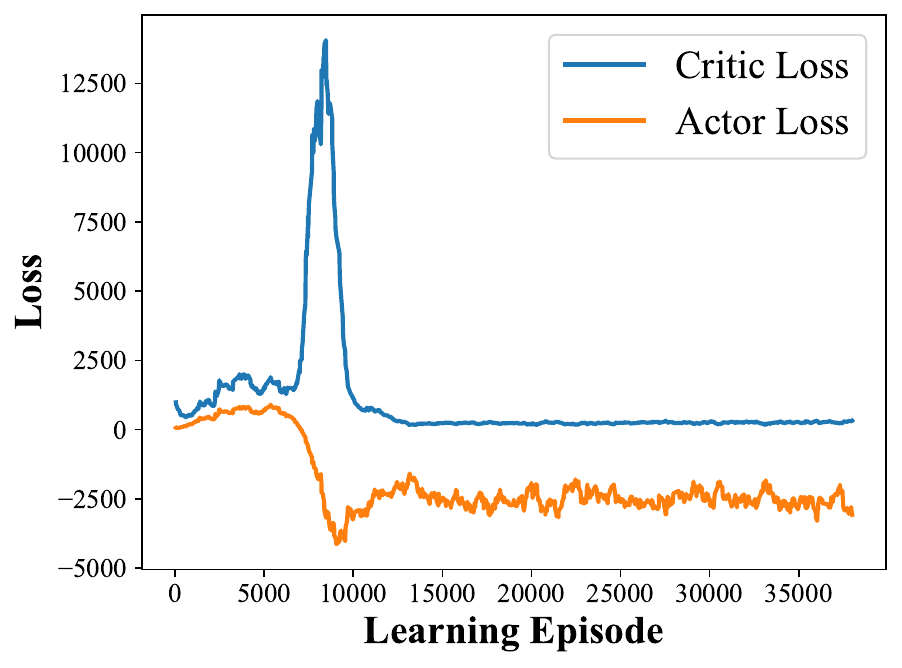}
        \label{fig_loss}
    \end{minipage}
    \caption{The values of reward, actor loss, and critic loss in the training process of the DTDRLMO.}
    \label{fig_training}
\end{figure}

\textbf{Configuration \& Baselines.} DTDRLMO is implemented on our platform equipped with $32$ cores and $2$ GPU cards. As presented in Fig. \ref{fig_training}, the training and convergence results of the DTDRLMO include the reward, actor loss, and critic loss. Our experimental results show that the training converges after approximately $8000$ episodes. The critic loss approximates to zero after $15000$ learning episodes, where each learning episode corresponds to an iteration of the learning function that updates the actor and critic networks. The training parameters are as follows: the replay memory size is set to $10000$, with a mini-batch size of $64$, indicating the number of memories used for each training step. The learning rates of actor and critic networks are set at $0.001$ and $0.002$, respectively. The reward discount is set at $0.9$, and we use a soft replacement parameter of $0.01$. We implement four comparative algorithms for microservice offloading to benchmark performance. The baselines are as follows:

\begin{itemize}
    \item Local Execution (LE): In this strategy, microservices are run on end devices, negating the need for network flow scheduling.
    
    \item Greedy: This strategy computes the estimated completion time for each device and prefers the device with the minimal time for offloading. Network flow scheduling follows the First Come First Served (FCFS) strategy.
    
    \item DRL without DT: Wang et al. \cite{wang2021computation} introduced a DRL-based computation offloading algorithm based on DDPG, albeit without adopting the digital twin technology.
\end{itemize}

\begin{figure*}[t]
    \centering
    \begin{minipage}[t]{0.27\linewidth}
        \centering
        \includegraphics[width = 0.83\linewidth]{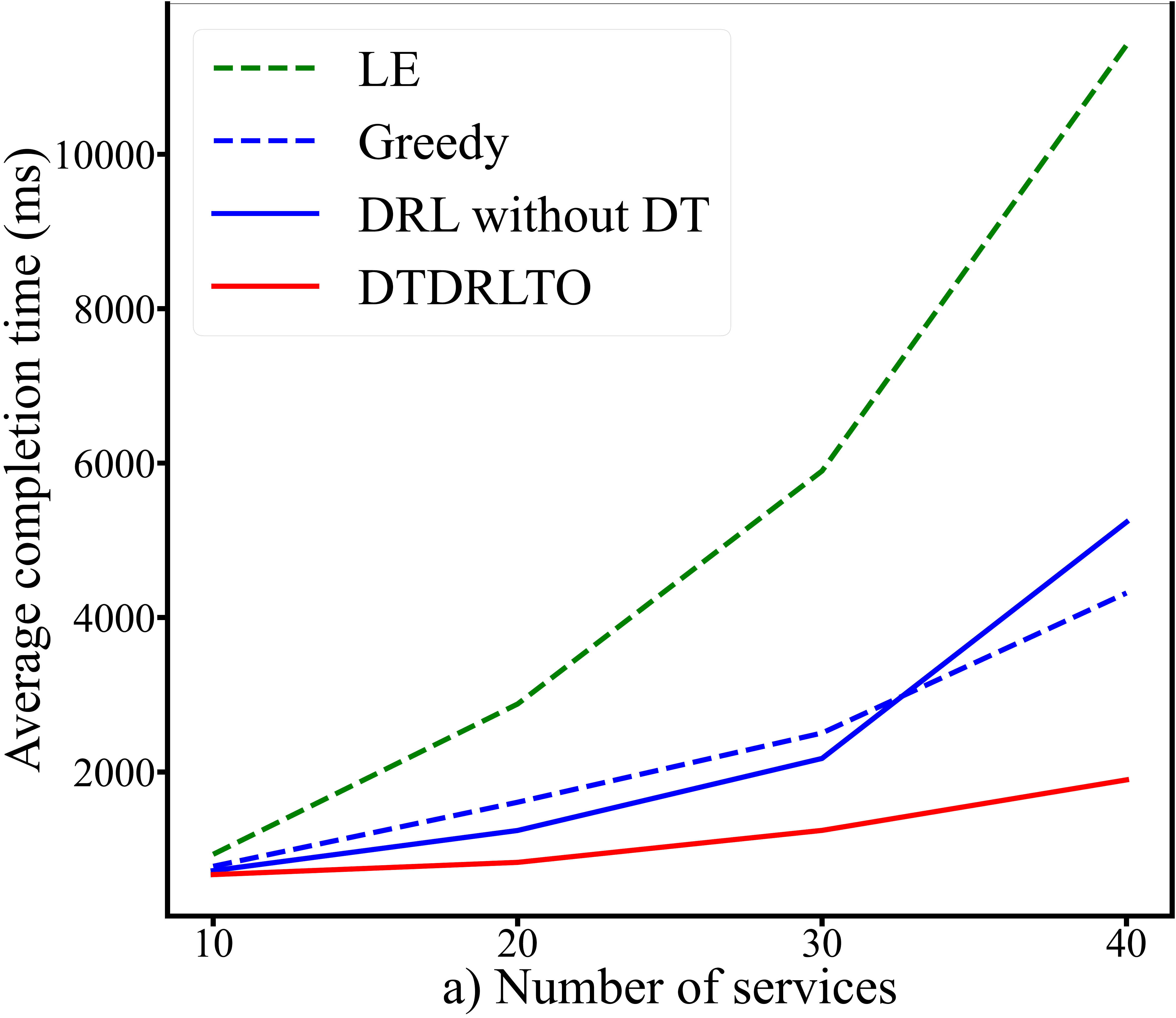}
        \label{fig_task}
    \end{minipage} 
    \begin{minipage}[t]{0.28\linewidth}
        \centering
        \includegraphics[width = 0.83\linewidth]{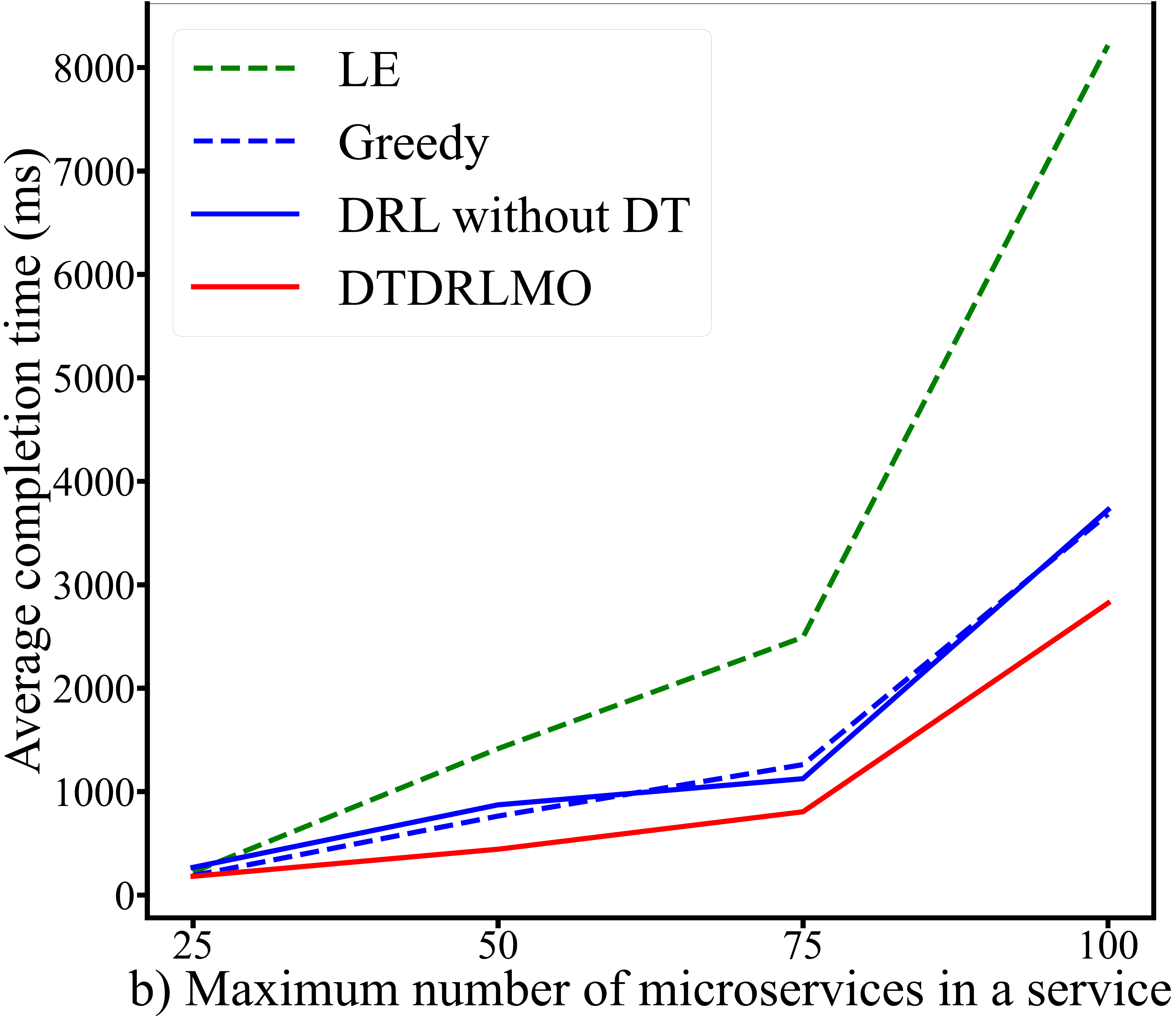}
        \label{fig_microservice}
    \end{minipage}%
    \begin{minipage}[t]{0.27\linewidth}
        \centering
        \includegraphics[width = 0.83\linewidth]{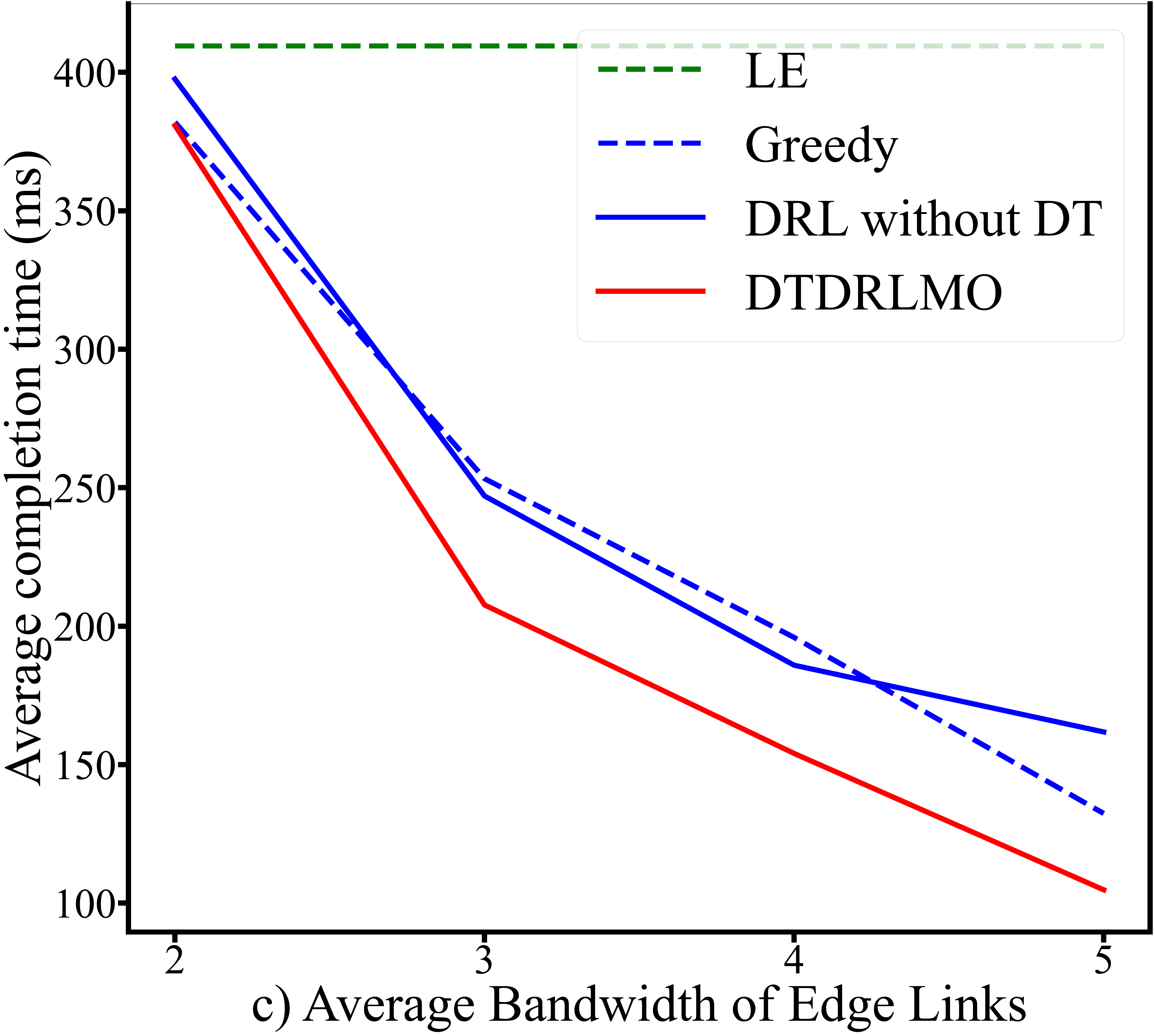}
        \label{fig_bandwidth}
    \end{minipage}%
    
    \caption{Simulation Results for Real-world Dataset}
    \label{fig_ali}
\end{figure*}

\begin{figure*}[t]
    \centering

    \begin{minipage}{0.26\linewidth}
        \centering
        \includegraphics[width= 0.83\linewidth]{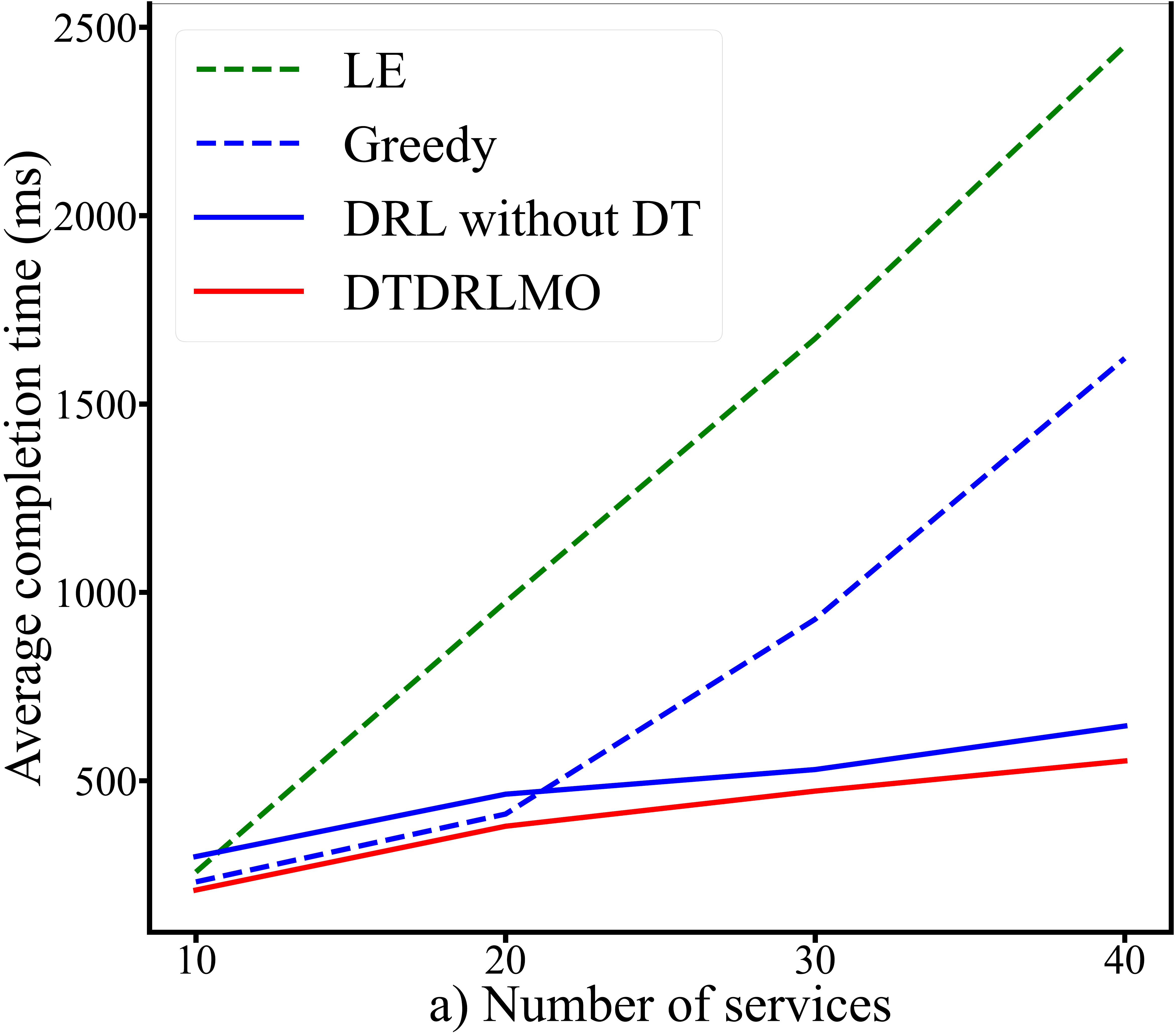}
        \label{fig_rtask}
    \end{minipage}
    \begin{minipage}{0.28\linewidth}
        \centering
        \includegraphics[width = 0.83\linewidth]{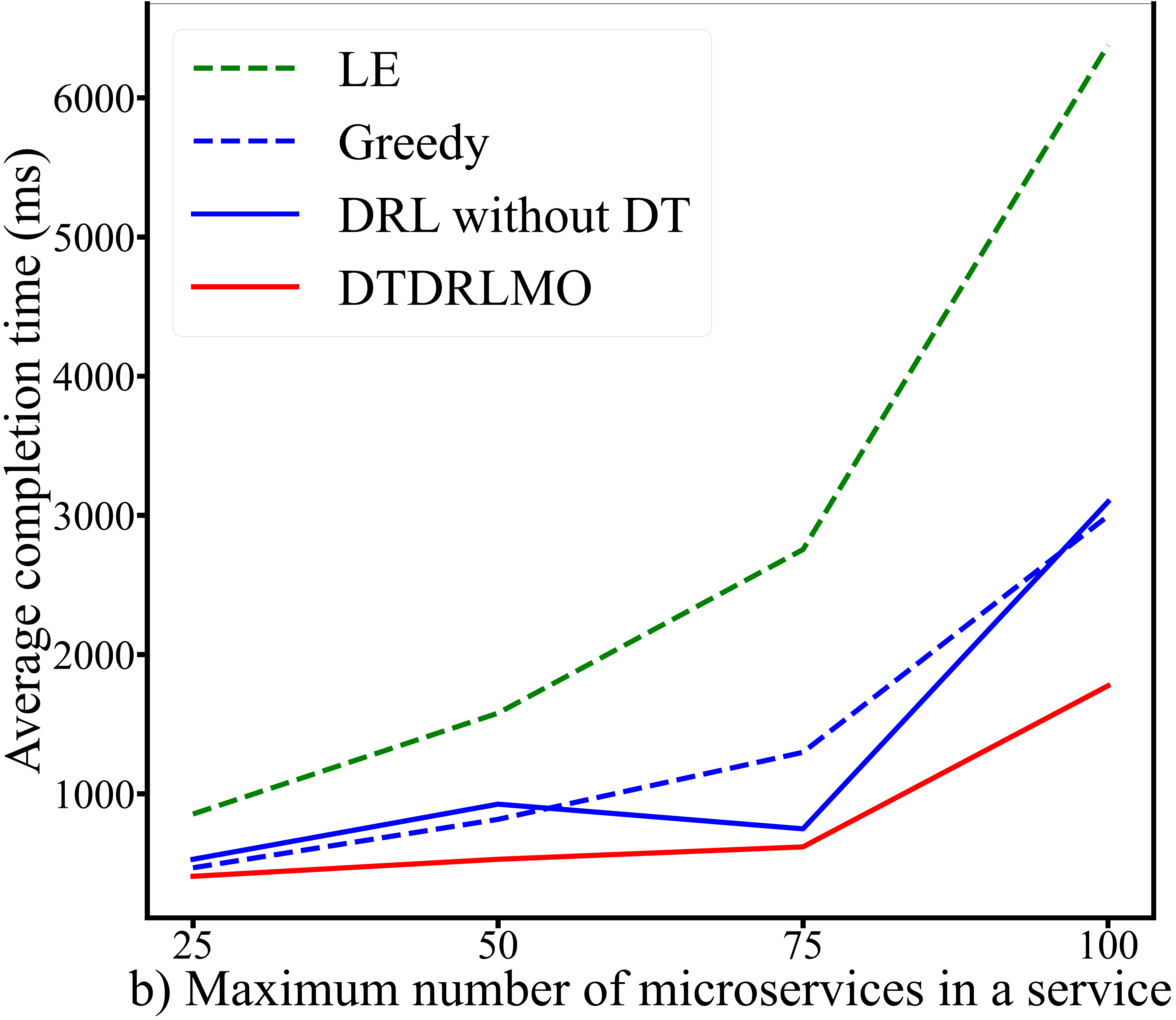}
        \label{fig_rmicroservice}
    \end{minipage}%
    \vspace{0.01cm}
    \begin{minipage}{0.26\linewidth}
        \centering
        \includegraphics[width = 0.83\linewidth]{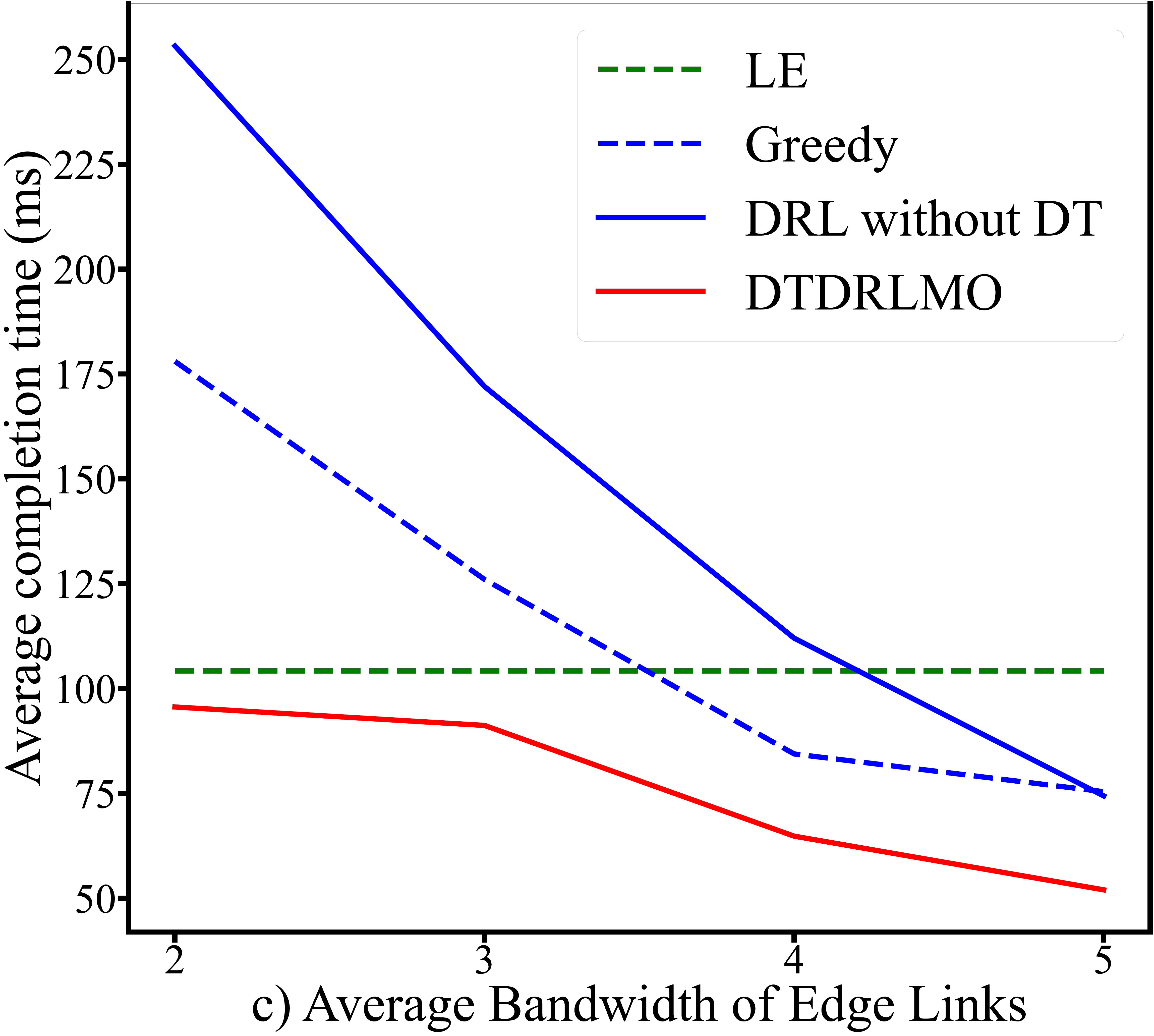}
        \label{fig_rbandwidth}
    \end{minipage}%
    
    \caption{Simulation Results for Synthetic Dataset}
    \label{fig_syn}
\end{figure*}

\subsection{Performance in the Real-world Dataset}

\textbf{Effect of Change in the Number of Services:} Fig.~\ref{fig_ali}(a) shows the effect of changing the number of services from $10$ to $40$ on ACT. The ACT difference between LE and DTDRLMO increases from $28.1\%$ at $10$ services to $83.4\%$ at $40$ services. This increase in the gap is observed because LE executes the microservices locally at the devices at which it is generated, which leads to a larger waiting time as the number of services is increased. When the number of services increases to $40$, DTDRLMO achieves an average delay of $59.9\%$ lower than those of DRL without DT and Greedy. This is because DTDRLMO, unlike Greedy and DRL without DT, incorporates digital twin technology, which allows it to better adapt to changes in the network environment and make more efficient offloading decisions. 

\textbf{Effect of Change in the Number of Microservices:} Fig.~\ref{fig_ali}(b) shows the effect of changing the maximum number of microservices in a service DAG from $25$ to $100$ within each service DAG. LE executes the services locally at the devices at which it is generated, which leads to a larger waiting time as the number of microservices is increased. So there is a significant difference in ACT between DTDRLMO and LE from about $18.1\%$ at $25$ maximum microservices to $65.6\%$ at $100$ maximum microservices. The difference in ACT between DTDRLMO and DRL without DT is $24.2\%$ at $100$ maximum microservices. This is because DRL without DT lacks the ability to adapt to dynamic changes in the network environment due to the absence of digital twin technology. DTDRLMO is around $26.3\%$ lower than Greedy in ACT. This is because the Greedy strategy only considers immediate benefits without taking into account the overall network condition.

\textbf{Effect of Change in the Average Bandwidth of Edge Links:} In Fig.~\ref{fig_ali}(c), the average bandwidth of edge links changes from $2$ Mbps to $5$ Mbps. The ACT of LE is constant due to the absence of network flows. It is expected that DTDRLMO performs significantly better than baselines in low bandwidth condition. When the average bandwidth decreases to $2$ Mbps, DTDRLMO achieves an ACT that is around $18.0\%$ and $15.9\%$ lower than that of DRL without DT and Greedy, respectively. This is because DRL without DT, lacking the digital twin technology, cannot adapt to the low bandwidth condition and thus makes inefficient offloading decisions, leading to a higher ACT. On the other hand, Greedy is not able to make full use of computation and bandwidth resources when offloading microservices to target nodes. 

\subsection{Performance in the Synthetic DataSet}

\textbf{Effect of Change in the Number of Services:} Fig.~\ref{fig_syn}(a) shows an increase in ACT for all algorithms of changing the number of services from $10$ to $40$. This is due to both increases in waiting time at the devices to execute the services and the total number of network flows. DTDRLMO is significantly better than LE. Because LE executes the service at the local device where the service is generated, resulting in an increase in waiting time for microservice execution. DTDRLMO's superiority is further evidenced as DRL without DT is around $22.4\%$ higher than the DTDRLMO at microservice $20$. This is because DRL without DT lacks the ability to adapt to dynamic changes in the network environment, leading to inefficient offloading decisions and consequently higher ACT. Furthermore, DTDRLMO averaged about $33.0\%$ lower than Greedy in ACT when the number of services ranged from $10$ to $40$. This is because Greedy often leads to suboptimal offloading decisions without taking into account the overall network condition.

\textbf{Effect of Change in the Number of Microservices:} In Fig.~\ref{fig_ali}(b), the maximum number of microservices in a service DAG changes from $25$ to $100$ within each service DAG. DTDRLMO is significantly better than LE. This is observed because LE does not offload microservices, and the computational load on the edge nodes increases with the number of microservices, resulting in longer waiting times and hence a higher ACT. DTDRLMO, on the other hand, obtains around $35.1\%$ higher performance than Greedy when the number of microservices increases to $50$. This is because DTDRLMO, unlike the Greedy strategy, is capable of making more informed offloading decisions by considering the overall network condition rather than just the immediate completion time of each edge node. When the number of microservices increases to $100$, DTDRLMO achieves an average delay of $41.7\%$ lower than those of DRL without DT and Greedy. Because DRL without DT and the Greedy strategy, due to their lack of adaptability and short-sighted nature in dynamic environment respectively, are unable to efficiently manage the increased computational load, leading to a higher delay.

\textbf{Effect of Change in the Average Bandwidth of Edge Links:} In Fig.~\ref{fig_syn}(c), the average bandwidth of edge links changes from $2$ Mbps to $5$ Mbps. The ACT of LE is constant due to all microservices run on local edge nodes where microservices are generated, and network flow scheduling is ignored. When the average bandwidth increases to $5$ Mbps, DTDRLMO achieves a delay of $50.1\%$ lower than LE. This is because DTDRLMO, unlike LE, leverages the increased bandwidth to offload microservices more efficiently. The ACT of Greedy and DRL without DT is significantly higher than LE at low bandwidth of $2$ Mbps. This is because both Greedy and DRL without DT rely on network bandwidth for offloading microservices, and their performance deteriorates when the bandwidth is limited. There is a decrease in ACT performance difference between DTDRLMO and DRL without DT from about $66.2\%$ at the average bandwidth of $2$ Mbps to $31.0\%$ at the average bandwidth of $5$ Mbps. This is because DRL without DT, despite its limitations, can still benefit from the increased bandwidth to some extent. DTDRLMO is around $87.5\%$ lower than Greedy in terms of ACT at the average bandwidth of $2$ Mbps and $31.0\%$ less at the average bandwidth of $5$ Mbps. This is because DTDRLMO, with its ability to learn and adapt, can make more efficient use of the available bandwidth compared to the short-sighted Greedy algorithm.

\section{Conclusion}\label{sec:conclusion}

In this paper, we formulate an online joint microservice offloading and bandwidth allocation problem, which addresses the dynamic resource availability issues of microservice offloading in CEC. We propose a novel algorithm, DTDRLMO, that combines DRL and digital twin techniques, enabling real-time adaptation to changing load on edge nodes and network conditions. Simulation results on real-world and synthetic datasets highlight that DTDRLMO outperforms state-of-the-art heuristic and learning-based methods. Our future work will implement and test our algorithm in a real-world edge computing system to further validate its effectiveness.

\section{Acknowledgement}
This work was supported by the Research Institute for Artificial Intelligence of Things, The Hong Kong Polytechnic University, the Hong Kong (HK) Research Grant Council (RGC) General Research Fund with project code PolyU 15220922 and the HK RGC Research Impact Fund with project code R5060-19.


\bibliographystyle{ieeetr}

\end{document}